\documentclass[preprint]{aastex}
\usepackage{graphicx}
\shorttitle{VB 10 Radial Velocities}
\shortauthors{Anglada-Escud\acute{e} et al.}

\begin{document}

%\title{New Constraints to the Orbit of the Planet Candidate
%around VB 10 using Doppler spectroscopy
\title{Strong Constraints to the Putative Planet Candidate
around VB 10 using Doppler spectroscopy
\footnote{Based on observations collected with the 6.5 meter Magellan
Telescopes located at Las Campanas Observatory, Chile, at the W. M. Keck
Observatory and the Canada-France-Hawaii Telescope (CFHT). The Keck
Observatory is operated as a scientific partnership between the California
Institute of Technology, the University of California, and NASA, and was
made possible by the generous financial support of the W. M. Keck
Foundation. CFHT is operated by the National Research Council of Canada,
the Institut National des Sciences de l'Univers of the Centre National de
la Recherche Scientique of France, and the University of Hawaii. }
}

\author{Guillem Anglada-Escud\'{e}}
\affil{Department of Terrestrial Magnetism, Carnegie Institution of
Washington\\
5241 Broad Branch Road, NW, Washington, DC 20015 USA}
\email{anglada@dtm.ciw.edu}

\author{Evgenya Shkolnik}
\affil{Department of Terrestrial Magnetism, Carnegie Institution of
Washington\\
5241 Broad Branch Road, NW, Washington, DC 20015 USA}
\email{shkolnik@dtm.ciw.edu}

\author{Alycia J. Weinberger}
\affil{Department of Terrestrial Magnetism, Carnegie Institution of
Washington\\
5241 Broad Branch Road, NW, Washington, DC 20015 USA}
\email{weinberger@dtm.ciw.edu}

\author{Ian B. Thompson}
\affil{The Observatories of the Carnegie Institution of Washington\\
813 Santa Barbara Street, Pasadena, CA 91101 USA}
\email{ian@obs.carnegiescience.edu}

\author{David J. Osip}
\affil{Las Campanas Observatory, Carnegie Institution of Washington\\
Colina El Pino Casilla 601, La Serena, Chile}
\email{dosip@lco.cl}

\author{John H. Debes}
\affil{Goddard Space Flight Center, NASA Postdoctoral Program\\
8463 Greenbelt Rd, Greenbelt, MD 20770, USA}
\email{john.H.debes@nasa.gov}

\begin{abstract}
We present new radial velocity measurements of the ultra-cool dwarf
VB~10, which was recently announced to host a giant planet detected with
astrometry. The new observations were obtained using optical spectrographs(MIKE/Magellan and ESPaDOnS/CHFT) and cover a 63\% of the reported period of 270 days. We apply Least-squares periodograms to identify the most significant signals and evaluate their corresponding False Alarm Probabilities. We show that this method is the proper generalization to astrometric data because (1) it mitigates the coupling of the orbital parameters with the parallax and proper motion, and (2) it permits a direct generalization to include non-linear Keplerian parameters in a combined fit to astrometry and radial velocity data.  In fact,
our analysis of the astrometry alone uncovers the reported
270~d period and an even stronger signal at $\sim$50 days. We estimate
the uncertainties in the parameters using a Markov Chain Monte Carlo approach. The nominal precision of the new Doppler measurements is about $150$ s$^{-1}$ while their standard deviation is $250$ ms$^{-1}$. However, the best fit solutions still have RMS of $200$ ms$^{-1}$ indicating that the excess in variability is due to uncontrolled systematic errors rather than the candidate companions detected in the astrometry. Although the new data alone cannot rule-out the presence of a candidate, when combined with published radial velocity measurements, the False Alarm Probabilities of the best solutions grow to unacceptable levels strongly suggesting that the observed astrometric wobble is not due to an unseen companion.

\end{abstract}

\section{Introduction}

\citet{Pravdo:2009} recently announced the discovery of an astrometric
companion to VB 10, an ultra-cool dwarf with a mass of $\approx$0.08
M$_\odot$. From a Keplerian fit to the motion, they determined a mass of 6
M$_{\rm J}$ and a period of 270 d. Thus VB 10 became the lowest mass star
known to harbor a planetary companion.  The mass ratio between VB 10 and its
companion, $\sim$13, also is intriguing. A similar mass ratio for a Solar-type
star would make the companion a brown dwarf, but brown dwarfs as small
separation companions to stars are quite rare.  VB 10 is itself the secondary
in a wide binary with V1428 Aql, a M2.5 star \citep{Biesbroeck:1961}.  At a
distance of 5.8 pc from the Sun, the 74\arcsec\ separation of this proper
motion binary corresponds to a projected separation of 430 AU.

Such low-mass stars have not been the target of intensive precision
radial velocity (PRV) monitoring because they have low visual fluxes and
high stellar activity. For example, the dedicated HARPS M-dwarf planet
search observes
stars\footnote{http://www.eso.org/sci/observing/proposals/77/gto/harps/3.txt}
only brighter than V=14 and of moderate to low activity levels
\citep{bonf07}. VB 10 has V mag = 17.3 and is known to be a flare star
\citep{berg08}. PRV and lensing planet searches have so far found only
13 stars under 0.5 M$_\odot$ hosting 18 planets, and of these, more than
half have masses below 0.1 M$_{\rm J}$.

Despite the challenges, searches for planetary companions to low mass
stars are of continuing interest. Low-mass stars appear less likely to
have lower mass stellar companions and less likely to harbor planets than
Solar-mass stars \citep{Cumming:2008}.  When they do have companions, they
tend to be stars of nearly equal mass to the primary
\citep{Burgasser:2007}.  The mass function of planets orbiting M dwarfs,
and how it differs from the planet mass function for higher-mass stars,
provides a constraint on the planet formation mechanism(s) in general.
Disks sufficiently massive to form Jupiter-mass planets appear to be rare
around brown dwarfs, whose disks generally look like lower mass versions
of T Tauri disks \citep{Scholz:2006}.  High mass companions would have to
form via a binary-like fragmentation mechanism
\cite[e.g.][]{Font-Ribera:2009}.  Thus how a 6 M$_{\rm J}$ planet could
form around an $\sim$80 M$_{\rm J}$ star and how common such high-mass
ratio companions remain important questions \citep{Boss:2009}.

The reported planet's astrometric orbit predicts a radial velocity (RV)
amplitude of at least 1 km s$^{-1}$ for a circular orbit and up to
several km s$^{-1}$ for an eccentric orbit.  This magnitude signal is
detectable with ordinary RV measurements without requiring the adoption
of precision techniques such as an iodine cell or simultaneous thorium
reference.

Although several RV measurements of VB 10 exist in the literature before
2009, it is difficult to combine the historical RVs (see list in Table 4
of \citet{Pravdo:2009}), as each observation used different calibration
techniques and/or RV standards that introduce zero-point offsets and
the typical uncertainties are also large ($\sim$1.5 km~s$^{-1}$). The most precise measurements in the literature were recently
published by \citet{Osorio:2009} (hereafter Z09), but provide only a
``hint of variability.'' These data did little to constrain the orbital
parameters of the planet beyond what the astrometry had already done
\citep{anglada:2009}.

Here, we present a more precise set of RV observations over 175 days (or 65\% of the reported orbital period). We also present general techniques for joint fitting of astrometric and RV data and show how they can be used to constrain the orbit of the candidate planet.

\section{New Data}\label{sec:spectra}

We acquired spectra at 7 epochs in 2009 with the MIKE spectrograph at the Magellan Clay telescope at Las Campanas Observatory (Chile). We used the 0.35$\arcsec$ and the 0.5$\arcsec$ slits  which produce a spectral resolution of $\approx$45,000 and 35,000, respectively, across the 4900 --
10000\AA\/  range of the red chip. The seeing was in the range from
$0.5$ to $1.1^{\prime\prime}$. These data were reduced using the
facility pipeline \citep{Kelson:2003}.

We also have in hand a single spectrum of VB 10 taken in 2006 using the
HIRES \citep{vogt94}) on the Keck I 10-m telescope. We used the 0.861$\arcsec$ slit to obtain a spectral resolution of $\lambda$/$\Delta\lambda$$\approx$58,000 at $\lambda\sim7000\AA$. We used the GG475 order-blocking filter and the red cross-disperser to maximize throughput in the red orders.

To increase the phase coverage, an additional spectrum was obtained
using the ESPaDOnS on the CFHT 3.6-m
telescope. ESPaDOnS is fiber fed from the Cassegrain to Coud\'e focus
where the fiber image is projected onto a Bowen-Walraven slicer at the
spectrograph entrance. ESPaDOnS' `star+sky' mode records the full
spectrum over 40 grating orders covering 3700 to 10400 \AA\/ at a
spectral resolution of $\lambda$/$\Delta\lambda$$\approx$68,000. The
data were reduced using {\it Libre Esprit} described in \cite{dona97,dona07}.

Each stellar exposure is bias-subtracted and flat-fielded for
pixel-to-pixel sensitivity variations. After optimal extraction, the 1-D
spectra are wavelength calibrated with a thorium-argon arc. To correct
for instrumental drifts, we used the telluric molecular oxygen A band
(from 7620 -- 7660 \AA) which aligns the MIKE spectra to 40 m~s$^{-1}$,
after which we corrected for the heliocentric velocity. Consistency
tests with the bluer Oxygen band shows comparable values
but with larger measurement error.

The final spectra
are of moderate S/N reaching $\approx$ 25 per pixel at 8000 \AA. Each
night, spectra were also taken of a M-dwarf RV standard, namely GJ 699
(Barnard's star; SpT = M4V) and/or GJ 908 (SpT = M1V).

To measure VB~10's RV, we cross-correlated each of 9 orders between 7000
and 9000~\AA\/ (excluding those with strong telluric absorption) where VB
10 emits most of its optical light, with the spectrum of GJ 699 and/or GJ
908 taken on the same night using IRAF's\footnote[1]{IRAF (Image
Reduction and Analysis Facility), http://iraf.noao.edu/} {\it fxcor} routine
(\citealt{fitz93}).  Both GJ 699 and GJ 908 have been monitored for
planets for years and none has been found within the RV stability level
of 0.1 km~s$^{-1}$.  Here we use the systemic RVs published by a
planet-search team \citep{nide02}: RV(GJ 699) = --110.506 km~s$^{-1}$ and
RV(GJ 908) = --71.147 km~s$^{-1}$. The zero-point of the absolute RVs is
uncertain at the 0.4 km~s$^{-1}$ level. We measured the RVs from the
gaussian peak fitted to the cross-correlation function (CCF) of each
order and adopt the average RV of all orders with a mean standard
deviation of the individual measurements of 0.150~km~s$^{-1}$. The
average of all our measurements is 36.02 km~s$^{-1}$ with a standard
deviation of 0.25 km~s$^{-1}$.  An observing log with the measured RVs
and uncertainties for VB 10 is shown in Table~\ref{table_log}.

\section{Data Analysis: Combining Astrometry and Radial Velocities}

In this section, we reanalyze the original astrometric data to calculate
the likelihood of astrometrically allowed solutions, and then combine the
astrometry and RV data sets in a consistent framework to quantify how the
new RV measurements constrain the possible orbits of the candidate signals
observed in the astrometry of VB~10b.

\subsection{Least-squares periodograms}

The most popular method to look for periodicities in data is the
so-called Lomb-Scargle periodogram. A version adapted to deal with
astrometric two-dimensional data developed by \citet{Catanzarite:2006}
(Joint Lomb Scargle periodogram) was implemented in the discovery paper
of VB 10b \citep{Pravdo:2009}. Any method based on the Lomb-Scargle
periodogram performs optimally only under an important implicit
assumption: all other signals (e.g. linear trend, an average offset,
etc.) can be subtracted from the data without affecting the significance
of the signal under investigation. This assumption does not hold for
astrometry because the proper motion and the parallax are also a
significant part of the signal and they typically correlate with the
periodic motion of a planet (see \citealt{black:1982}).

We use instead a Least-squares periodogram. The weighted Least-squares
solution is obtained by fitting all the free parameters in the model for a
given period. The sum of the weighted residuals divided by $N$ is the
so-called $\chi^2$ statistic. Then, each $\chi^2_P$ of a given model with
$k_P$ parameters, can be compared to the $\chi^2_0$ of the null hypothesis
with $k_0$ free parameters by computing the power, $z$, as
\begin{eqnarray}
z(P) = \frac{(\chi^2_0-\chi^2_P)/(k_P-k_0)}{\chi^2_P/(N_{\rm obs}-k_P)}
\end{eqnarray}
\noindent where a large $z$ is interpreted as a very significant solution.
The values of $z$ follow a Fisher F-distribution with $k_P-k_0$ and
$N_{\rm obs}-k_P$ degrees of freedom \citep{Scargle:1982,Cumming:2004}.
Even if only noise is present, a periodogram will contain several peaks
\citep[see][as an example]{Scargle:1982} whose existence have to be
considered in obtaining the probability of a spurious detection. Assuming
Gaussian noise, the probability that a peak in the periodogram has a power
higher than $z(P)$ by chance is the so-called False Alarm Probability
(FAP) :
\begin{eqnarray}
{\rm FAP} = 1-\left(1-{\rm Prob}[z>z(P)]\right)^M \label{eq:faps}
\end{eqnarray}
\noindent where $M$ is the number of independent frequencies. In the case
of uneven sampling, $M$ can be quite large and is roughly the number of
periodogram peaks one could expect from a data set with only Gaussian
noise and the same cadence as the real observations. We adopt the recipe
$M\approx 2\Delta T/P_{min}$ given in \citet[][Sec 2.2]{Cumming:2004},
where $\Delta T$ is the time-span of the observations and $P_{\rm min}$ is
the minimum period searched. One still has to select $P_{\rm min}$
arbitrarily. Assuming a $P_{\rm min}=20$ days, the astrometric data alone
has $M\sim 300$, and the combination of astrometry and RVs has $M \sim
360$.

In our particular problem, the null hypothesis is the basic kinematic
model with $k_0=6$ parameters: 2 coordinates, 2 proper motions, parallax
and systemic RV. As a first approach, our simplest non-null hypothesis
considers circular orbits, astrometric data only and one RV measurement.
For a given period, the number of free parameters is then $k_P = 10$:
the 6 kinematic ones plus the four Thiele Innes elements $A$, $B$, $F$
and $G$ \citep[e.g.][]{wright:2009}. Since the model is linear in all
$10$ parameters, the power can be efficiently computed for many periods
between $20$ days and $4000$ days to obtain a familiar representation of
the periodogram that we call a \textit{Circular Least-squares
Periodogram} (CLP). The CLP of the astrometric data, shown at top in
Figure~\ref{fig:periodogram}, displays two obvious peaks: the reported
one at 270 days \citep{Pravdo:2009} and a more significant one at 49.9
days, both with high power and very low FAPs.

To find the full Keplerian solution for both periods and estimate their
FAPs, we perform a Least-squares periodogram sampling a grid of fixed
eccentricity-period (eP) pairs and fitting all other parameters. For
each eP pair $k_P$ is $11$: the null-hypothesis ($X_0$, $Y_0$,  $\mu_X$,
$\mu_Y$, $\pi$ and $v_0$) plus all the other Keplerian elements: Mass of
the planet, $\Omega$, $\omega$, $i$, and the Mean anomaly at the initial
epoch $M_0$ \citep[see][for a recent review]{wright:2009}. We analyze
both \textit{astrometry only} and \textit{astrometry+RVs}. The $\chi^2$
of the best fit solution is then used to obtain each FAP as previously
described. Figure \ref{fig:periodogram} shows the resulting color-coded
FAPs for each eccentricity--period pair (eP-map).

\subsubsection{Astrometry only}

A value of $M=300$ has been used to obtain the FAP, and our result at
270~d qualitatively agrees with \citet{Pravdo:2009}, however the more
significant period is at $\sim$50~d. For both periods, there are regions
with FAP$<$1\% spanning all possible eccentricities (second row in
Figure \ref{fig:periodogram}). The best fits and their $\chi^2$ per degree
of freedom ($\bar{\chi}^2$) are summarized in Table~\ref{tab:pars}. The obtained results for the 270~d period are in agreement with those reported in the discovery paper by \citet{Pravdo:2009}. The best fit solution for the 50~d period has mass $\sim15$ M$_J$, which would be a very low mass brown dwarf. It is important to point out that the best fit inclination is close to $90$ (edge on) for both solutions. The uncertainties on the orbital parameters are quantified in Sec~\ref{sec:MCMC}.

\subsubsection{Astrometry+RVs}

\begin{figure}
%\plotone{fig2.eps}
\includegraphics[width=6in,clip]{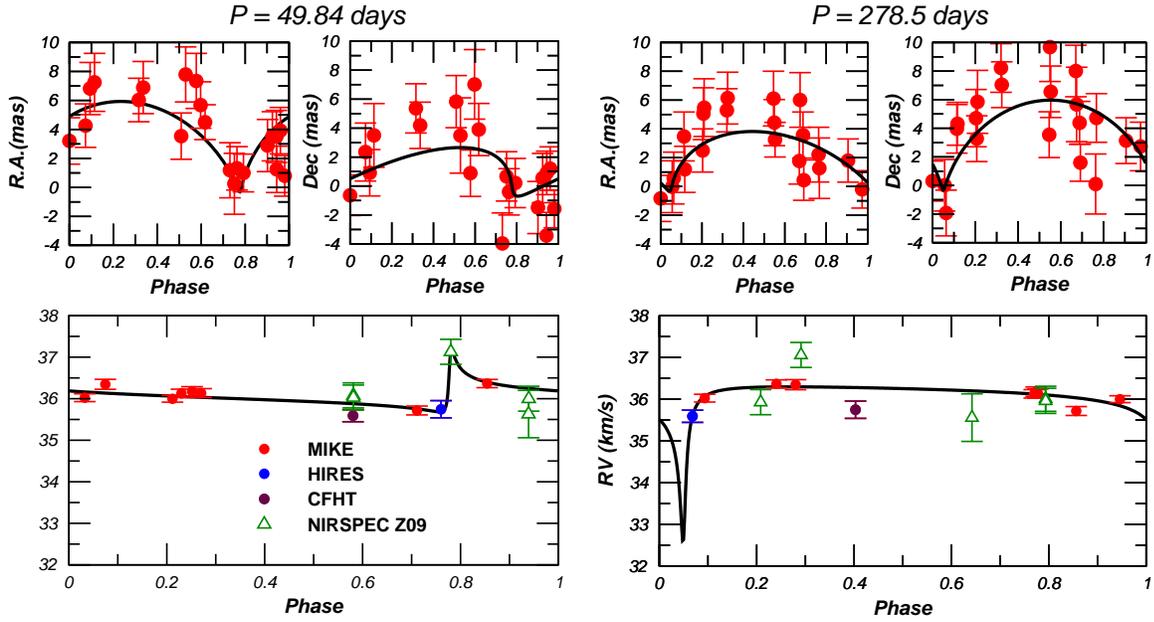}
\caption{The best fit (lowest $\chi^2$) joint solutions to the \citet{Pravdo:2009}
  astrometry and used RVs for the two signals.
  Top panels contain the astrometric offsets after the removal of the corresponding parallax and the proper motion.
  The lower panels contain all RVs used. Each RV point represents
  the weighted average of the values obtained using both reference stars if
  available. The best fit
  offset has been applied to Z09 data (Green triangles).
  Phase $0$ corresponds to the first astrometric epoch at JD
  $2451438.64$ \label{fig:higheccfit} and the corresponding folding periods
  are given on the top.}

\end{figure}

We now fit jointly for the best orbital solution to the astrometry
and RVs.  Our campaign covered about 65\%
of the 270~d orbit. The standard deviation of all our RVs measurements
is $250$ m s$^{-1}$ (null hypothesis) which is larger than the
individual uncertainties in Table \ref{tab:data}. When we
cross-correlate our standards, we measure a similar RMS of 200 m
s$^{-1}$, which indicates that the difference is due to an uncontrolled
or unmeasured systematic. The RMS of the RVs for the best fit solution is $200$ m s$^{-1}$, which is not statistically different from the RMS of the null hypothesis. This is another indication that our measurements contain systematic errors at the level of $100-200$ m/s. Despite of that, we use the nominal errors in the Least-squares solution as the best estimates for the individual uncertainties we can provide.  In Figure \ref{fig:higheccfit}, we show the best solutions to both signals including all the data.

\begin{figure}
\includegraphics[width=6in,clip]{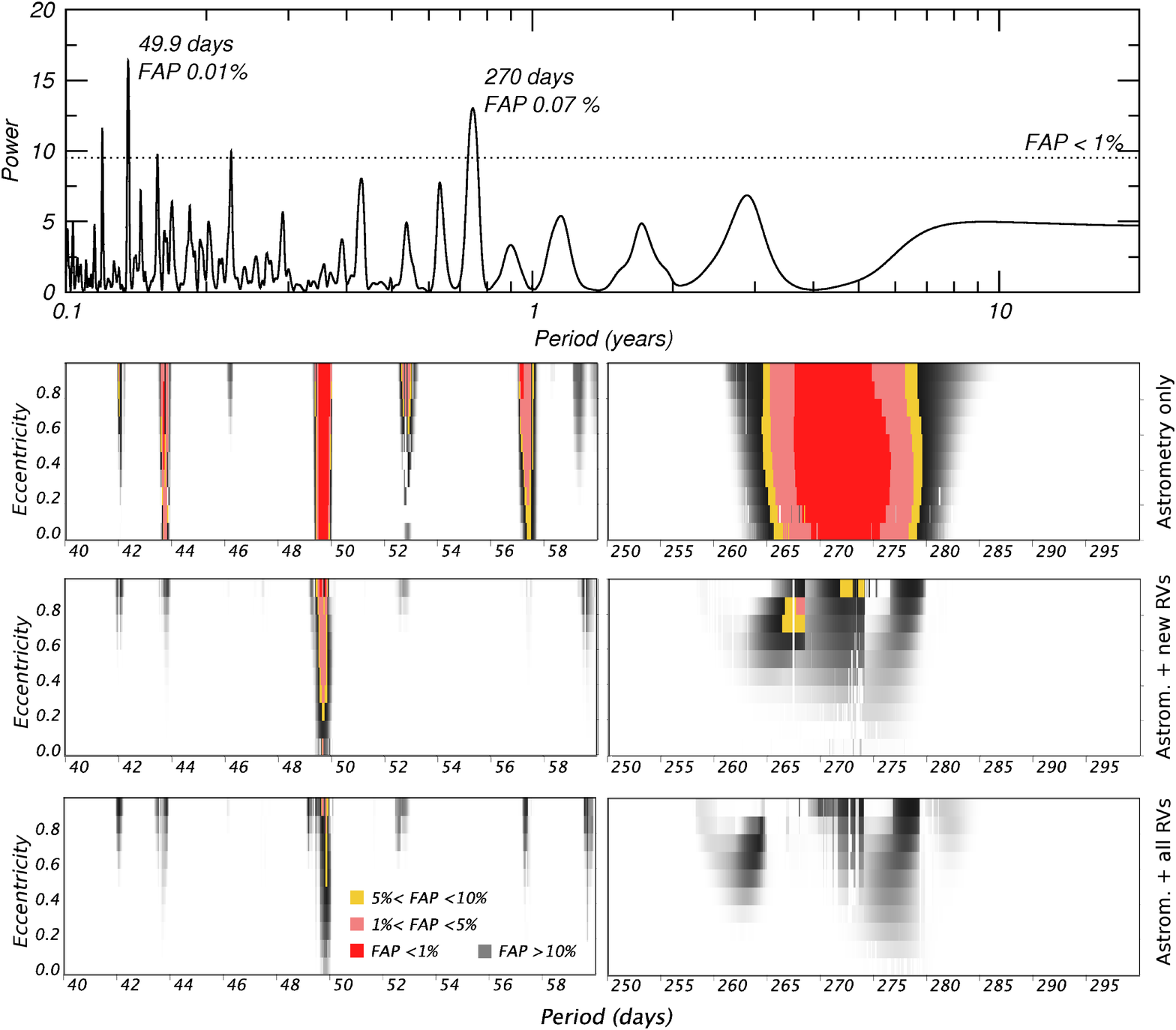}

\caption{\textbf{Top panel.} Circular Least-squares periodogram showing the two
most significant periods with their corresponding False Alarm Probabilities
(FAP). \textbf{Second row.} FAPs obtained for a grid of Eccentricity--Period
pairs around the 50~d (left) and the 270~d (right) when only astrometry is
considered.
\textbf{Third row.} FAPs obtained when our
new RV are included to the fit.
\textbf{Bottom row.} Final FAPs obtained when all published RV data are
combined in a joint fit.
\label{fig:periodogram}}
\end{figure}

For the 270~d period, our RV \textit{non detection} cannot exclude a small region of orbital solutions around $e\sim 0.8$ with a FAP between 1\%--5\% -- see Figure~\ref{fig:periodogram}, third row right panel.  We now add the RVs measurements by Z09 and solve for a joint solution. A zero-point offset between datasets is added as an additional free parameter. The combined
RV measurements force the eccentricity to large values which apparently still provides a reasonable fit to the astrometry (see top panels in Figure
\ref{fig:higheccfit}). However, the FAPs are now all higher than
$10$\% (Figure~\ref{fig:periodogram}, bottom right panel), which indicates that the signal can be barely distinguished
from the noise fluctuations. The ``hint'' of detection in Z09 based on
one discrepant value at $3.1-\sigma$ out of five can be due to random
errors with a non negligible probability.

For the 50~d period, there are still several orbits that provide a decent fit to the combined astrometry and the new RV data with a FAP lower
than 1\%. These occupy a small space around the
best joint solution, with $e=0.90$ (see Figure \ref{fig:periodogram},
3rd row, left panel) and an inclination close to $0$.
Large eccentricity causes the duration of fast RV variation to be very
short (and difficult to catch); an inclination close to $0$ tends to
suppress any RVs signal. Such inclination is in apparent contradiction
with the one obtained using the astrometry alone ($\sim 90$ deg). The
reason is the following: while the new fit to the astrometry forced by
the RVs is much worse than the one obtained from the astrometry alone,
such a solution still represents an improvement compared to the null
hypothesis. Adding Z09 data to the fit increases the FAP of the most
likely solution to 2\%, an eccentricity of $0.91$ and the inclination close to 0 (see Table~\ref{tab:pars}). This suggests that
the signal at 50~d is also spurious, even though it has slightly better
chances of survival than the one at 270~d.

\subsection{\textit{A Posteriori} Probability Distributions}\label{MCMC}
\label{sec:MCMC}

\begin{figure}
%\plotone{fig3.eps}
\includegraphics[width=6in,clip]{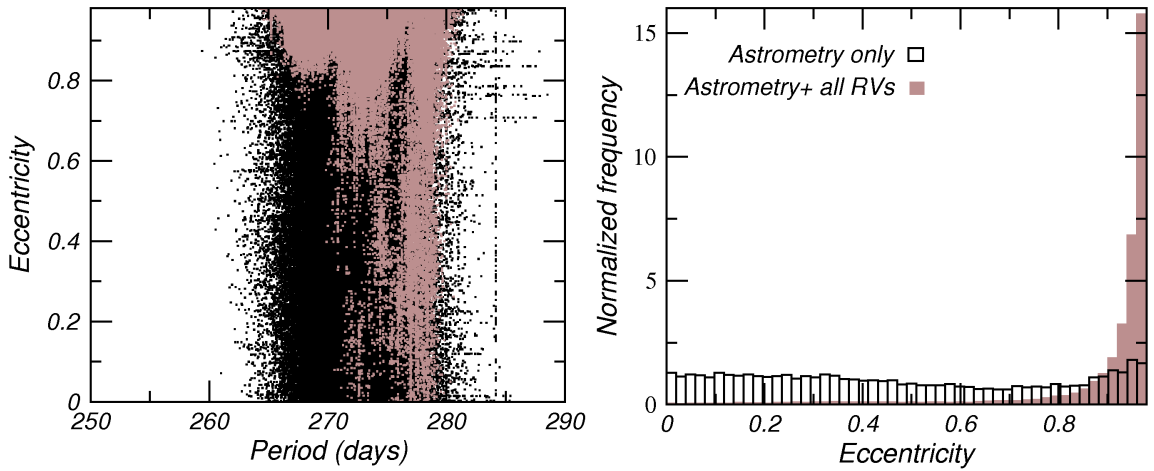}

\caption{
Left: Steps in period-eccentricity space of a Markov Chain of $10^6$
elements applied to the astrometry only (black) and to the astrometry+all
RV data (brown). The distribution resembles the FAP contours on
the eP-maps around $270$~days indicating that the chain has successfully
converged to the equilibrium distribution. Right: Histogram reproducing
the marginalized density distributions in $e$ for the astrometry only
and astrometry+RV around the 270~d solution.
\label{fig:dist}}
\end{figure}

We adapt the method developed by \citet{Ford:2005,Ford:2006} to assess
uncertainties in orbit determinations by obtaining the \textit{a
posteriori} probability distribution for the parameters using a Markov
Chain with a Gibbs sampler strategy. Our problem is identical to the
one described by \citet{Ford:2005}, where now the $\chi^2$ contains
both RV and astrometric observations and the model has a few more free
parameters. Several properly adjusted MCMC with $10^6$ steps have been
computed obtaining compatible results. The step sizes of the Gibbs
sampler are initialized with the formal errors from the best fit
Least-squares solution, and adjusted to obtain a transition
probability between 10\% and 20\%.  The first $10^5$ steps of each
chain are rejected. The final distributions match very well the areas
of low FAPs in the eP-maps (see Figure \ref{fig:dist} as an example)
giving further proof that the chains have converged to the equilibrium
distributions. The MCMC contains $13$ free parameters -- the $11$ from
the Least-squares periodogram plus eccentricity and period. When the RVs
measurements from Z09 are included, and additional offset parameter is
included.

Table \ref{tab:pars} presents the standard deviations obtained via the
MCMC for both the 50~d and 270~d periods using astrometry alone and
astrometry + all RV data. As an example, we show the two dimensional
density of states in period-eccentricity space in Figure \ref{fig:dist}
(left) obtained in both cases around the 270~d signal. The marginalized
distributions for $e$ in the form of histograms are shown in Figure
\ref{fig:dist} (right). For the astrometry-alone case, the distribution
of $e$ is almost uniform. It becomes strongly peaked towards high
eccentricities when all the RV data are included. Since the best fit
solution at 270~d is poor ($\bar{\chi}^2=1.76$), the corresponding
$\chi^2$ minimum is not very deep which is reflected in a significant
increase in the derived uncertainties (See Table \ref{tab:pars}). The
same happens to the signal at 50~d with the exception of the
inclination that has a small uncertainty ($4$ deg) close to 0. Even though this solution has a low FAP, the inclination has to be coincidentally very small to suppress any RV signal and very different from using astrometry only ($94$ deg), rasing serious doubts of its reality.

\section{Discussion and Conclusions}

The non-detection of a significant RV variation in our data set already
discards most orbital configurations allowed by the astrometry. When
combined with Z09 RVs measurements, there are no remaining solutions
with a FAP lower than 10\% around the 270~d period, so the presence of
a planet candidate at that period is not supported by the
observations. For the 50~d period, the constraints are also strong and
become almost definitive when the Z09 data is included. Even highly
eccentric solutions have a relatively large FAP ($>2\%$). We find that
particular combinations of eccentricity, inclination and $\omega$ can
fit an almost flat RV curve indicating that the analytic methods
applied to estimate FAPs for high eccentricities tend to give over
optimistic results and that this issue should be studied in more
detail.

We have developed and implemented useful tools for detailed analysis of
combined astrometric and RV data: Circular Least-squares periodogram as
the proper generalization of the classic Lomb-Scargle periodogram to
deal with astrometric data, eP-maps to visualize the most likely
period--eccentricity combinations and a Bayesian characterization of
the parameter uncertainties based on a MCMC approach.

VB~10 is also part of the Carnegie Astrometric Planet Search program
\citep{Boss:2009}. RV measurements with precision techniques in the
near-infrared \citep{Bean:2009} may provide the required accuracy to put
even stronger limits to the existence of VB10b or find other planets in
the system. VB~10 will certainly be observed by the space astrometry
mission Gaia \citep{Perryman:2001}, which would be capable of finding a
planet with a period of 270~d and as small as $0.2$ M$_J$.

\bibliographystyle{apj}

\begin{thebibliography}{28}
\expandafter\ifx\csname natexlab\endcsname\relax\def\natexlab#1{#1}\fi

\bibitem[{Anglada-Escud\'e {et~al.}(2009)Anglada-Escud\'e, Boss, \&
  Weinberger}]{anglada:2009}
Anglada-Escud\'e, G., Boss, A.~P., \& Weinberger, A.~J. 2009, in ASP Conf.
  Ser., Vol. in press, {Pathways Towards Habitable Planets}, ed. {V. Coud\'e du
  Foresto, D. M. Gelino \& I. Ribas} (San Francisco, CA: ASP)

\bibitem[{{Bean} {et~al.}(2009){Bean}, {Seifahrt}, {Hartman}, {Nilsson},
  {Wiedemann}, {Reiners}, {Dreizler}, \& {Henry}}]{Bean:2009}
{Bean}, J.~L., {Seifahrt}, A., {Hartman}, H., {Nilsson}, H., {Wiedemann}, G.,
  {Reiners}, A., {Dreizler}, S., \& {Henry}, T.~J. 2009, in ASP Conf. Ser.,
  Vol. in press, {Pathways Towards Habitable Planets}, ed. {V. Coud\'e du
  Foresto, D. M. Gelino \& I. Ribas} (San Francisco, CA: ASP), arXiv:0911.3148

\bibitem[{{Berger} {et~al.}(2008)}]{berg08}
{Berger}, E. {et~al.} 2008, \apj, 676, 1307

\bibitem[{Black \& Scargle(1982)}]{black:1982}
Black, D.~C. \& Scargle, J.~D. 1982, ApJ, 263, 854

\bibitem[{{Bonfils} {et~al.}(2007)}]{bonf07}
{Bonfils}, X. {et~al.} 2007, \aap, 474, 293

\bibitem[{{Boss} {et~al.}(2009)}]{Boss:2009}
{Boss}, A.~P. {et~al.} 2009, \pasp, 121, 1218

\bibitem[{{Burgasser} {et~al.}(2007){Burgasser}, {Reid}, {Siegler}, {Close},
  {Allen}, {Lowrance}, \& {Gizis}}]{Burgasser:2007}
{Burgasser}, A.~J., {Reid}, I.~N., {Siegler}, N., {Close}, L., {Allen}, P.,
  {Lowrance}, P., \& {Gizis}, J. 2007, in Protostars and Planets V, ed.
  {B.~Reipurth, D.~Jewitt, \& K.~Keil} (Tucson, AZ: Univ. of AZ Press), 427

\bibitem[{Catanzarite {et~al.}(2006)Catanzarite, Shao, Tanner, Unwin, \&
  Yu}]{Catanzarite:2006}
Catanzarite, J., Shao, M., Tanner, A., Unwin, S., \& Yu, J. 2006, PASP, 118,
  1319

\bibitem[{Cumming(2004)}]{Cumming:2004}
Cumming, A. 2004, MNRAS, 354, 1165

\bibitem[{Cumming {et~al.}(2008)Cumming, Butler, Marcy, Vogt, Wright, \&
  Fischer}]{Cumming:2008}
Cumming, A., Butler, R.~P., Marcy, G.~W., Vogt, S.~S., Wright, J.~T., \&
  Fischer, D.~A. 2008, PASP, 120, 531

\bibitem[{{Donati} {et~al.}(2007){Donati}, {Jardine}, {Gregory}, {Petit},
  {Bouvier}, {Dougados}, {M{\'e}nard}, {Cameron}, {Harries}, {Jeffers}, \&
  {Paletou}}]{dona07}
{Donati}, J.-F., {Jardine}, M.~M., {Gregory}, S.~G., {Petit}, P., {Bouvier},
  J., {Dougados}, C., {M{\'e}nard}, F., {Cameron}, A.~C., {Harries}, T.~J.,
  {Jeffers}, S.~V., \& {Paletou}, F. 2007, \mnras, 380, 1297

\bibitem[{{Donati} {et~al.}(1997){Donati}, {Semel}, {Carter}, {Rees}, \&
  {Collier Cameron}}]{dona97}
{Donati}, J.-F., {Semel}, M., {Carter}, B.~D., {Rees}, D.~E., \& {Collier
  Cameron}, A. 1997, \mnras, 291, 658


\bibitem[{{Fitzpatrick}(1993)}]{fitz93}
{Fitzpatrick}, M.~J. 1993, in ASP Conf. Ser., Vol.~52, Astronomical Data
  Analysis Software and Systems II, ed. R.~J. {Hanisch}, R.~J.~V. {Brissenden},
  \& J.~{Barnes}, 472

\bibitem[{{Font-Ribera} {et~al.}(2009){Font-Ribera}, {Miralda Escud{\'e}}, \&
  {Ribas}}]{Font-Ribera:2009}
{Font-Ribera}, A., {Miralda Escud{\'e}}, J., \& {Ribas}, I. 2009, \apj, 694,
  183

\bibitem[{{Ford}(2005)}]{Ford:2005}
{Ford}, E.~B. 2005, \aj, 129, 1706

\bibitem[{Ford(2006)}]{Ford:2006}
Ford, E.~B. 2006, ApJ, 642, 505

\bibitem[{Ford {et~al.}(2000)Ford, Kozinsky, \& Rasio}]{Ford:2000}
Ford, E.~B., Kozinsky, B., \& Rasio, F.~A. 2000, ApJ, 535, 385

\bibitem[{{Kelson}(2003)}]{Kelson:2003}
{Kelson}, D.~D. 2003, \pasp, 115, 688

\bibitem[{{Nidever} {et~al.}(2002){Nidever}, {Marcy}, {Butler}, {Fischer}, \&
  {Vogt}}]{nide02}
{Nidever}, D.~L., {Marcy}, G.~W., {Butler}, R.~P., {Fischer}, D.~A., \& {Vogt},
  S.~S. 2002, \apjs, 141, 503

\bibitem[{{Perryman} {et~al.}(2001)}]{Perryman:2001}
{Perryman}, M.~A.~C. {et~al.} 2001, \aap, 369, 339

\bibitem[{Pravdo \& Shaklan(2009)}]{Pravdo:2009}
Pravdo, S.~H. \& Shaklan, S.~B. 2009, ApJ, 700, 623

\bibitem[{{Scargle}(1982)}]{Scargle:1982}
{Scargle}, J.~D. 1982, \apj, 263, 835

\bibitem[{Scholz {et~al.}(2006)Scholz, Jayawardhana, , \& Wood}]{Scholz:2006}
Scholz, A., Jayawardhana, R., , \& Wood, K. 2006, ApJ, 645, 1498

\bibitem[{Tamuz {et~al.}(2008)}]{Tamuz:2008}
Tamuz, O. {et~al.} 2008, A\&A, 480, L33

\bibitem[{van Biesbroeck(1961)}]{Biesbroeck:1961}
van Biesbroeck, G. 1961, AJ, 66, 528

\bibitem[{{Vogt} {et~al.}(1994)}]{vogt94}
{Vogt}, S.~S. {et~al.} 1994, in SPIE Conf. Ser., Vol. 2198, Instrumentation in
  Astronomy VIII, ed. D.~L. {Crawford} \& E.~R. {Craine}, 362

\bibitem[{{Wright} \& {Howard}(2009)}]{wright:2009}
{Wright}, J.~T. \& {Howard}, A.~W. 2009, \apjs, 182, 205

\bibitem[{{Zapatero Osorio} {et~al.}(2009){Zapatero Osorio}, {Mart{\'{\i}}n},
  {del Burgo}, {Deshpande}, {Rodler}, \& {Montgomery}}]{Osorio:2009}
{Zapatero-Osorio}, M.~R., {Mart{\'{\i}}n}, E.~L., {del Burgo}, C., {Deshpande},
  R., {Rodler}, F., \& {Montgomery}, M.~M. 2009, \aap, 505, L5

\end{thebibliography}

\begin{deluxetable}{llcccrlllcccc}
\tabletypesize{\scriptsize}
%\rotate
\tablecaption{Log of RVs\label{table_log}}\label{tab:data}
\tablewidth{0pt}
\tablehead{
\colhead{Telescope} & \colhead{UT Date}  & \colhead{HJD}  &    \colhead{Slit Width} & \colhead{RV (w/ GJ 699)\tablenotemark{a}}  & \colhead{RV (w/ GJ 908)\tablenotemark{a}} \\
\colhead{+Instrument} & \colhead{} &  \colhead{--2450000} &    \colhead{$\arcsec$} & \colhead{km~s$^{-1}$} & \colhead{km~s$^{-1}$}
}
\startdata

%Telescope	&	Slit width	&	HJD	&	RV		sigma	&	RV STD
%+ Instrument	&	($\arcsec$	&	(-2450000)	&				&

Keck I + HIRES	&	2006 Aug 12     &	3959.57	&	     0.86    &       -- 		     &        35.59  $\pm$   0.15    \\
Clay+MIKE	&	2009 Jun 06	&	4988.74	&	     0.35    &       36.23   $\pm$    0.13   &       35.99   $\pm$   0.15    \\
Clay+MIKE	&	2009 Jun 07	&	4989.82	&	     0.50    &       36.22   $\pm$    0.13   &       36.09   $\pm$   0.20    \\
Clay+MIKE	&	2009 Jun 08	&	4990.75	&	     0.50    &       36.15   $\pm$    0.12   &       36.10   $\pm$   0.22    \\
Clay+MIKE	&	2009 Jun 30	&	5012.72	&	     0.35    &       35.72   $\pm$    0.11   &       -- 		     \\
Clay+MIKE	&	2009 Jul 25	&	5037.66	&	     0.50    &       35.96   $\pm$    0.11   &       36.03   $\pm$   0.11    \\
Clay+MIKE	&	2009 Sep 04	&	5078.58	&	     0.50    &       35.96   $\pm$    0.09   &       36.37   $\pm$   0.24    \\
Clay+MIKE	&	2009 Oct 15	&	5119.54	&	     0.50    &       36.30   $\pm$    0.14   &       36.41   $\pm$   0.13    \\
Clay+MIKE	&	2009 Oct 26	&	5130.51	&	     0.50    &       36.41   $\pm$    0.16   &       36.27   $\pm$   0.18    \\
CFHT+ESPaDOnS	&	2009 Nov 29	&	5164.69	&--\tablenotemark{b} &               --              &       35.74   $\pm$   0.20   \\

\enddata

\tablenotetext{a}{
  Uncertainties are the standard deviation of the 9
  orders of the cross correlation and do not include the
  40 m~s$^{-1}$ systematic uncertainty from the telluric
  wavelength correction. Absolute radial velocity
  determination has an uncertainty of $0.4$ km/s but it is not
  relevant for orbital fitting purposes.
}
\tablenotetext{b}{
  ESPaDOnS is a fiber fed spectrograph with an effective
  resolution of R$\sim$68000 in the wavelength range of interest
}
 % do not include the 40
%m~s$^{-1}$ systematic uncertainty from the telluric wavelength correction.}

\end{deluxetable}

\begin{deluxetable}{ccc|cc}
\tabletypesize{\scriptsize}
%\rotate

\tablecaption{Best fitting values\tablenotemark{a}. Uncertainties obtained from a MCMC with
$10^6$ steps\label{tab:pars}.}

\tablewidth{0pt}
\tablehead{
\colhead{Parameter} &
\colhead{Astrometry 50 d}  &
\colhead{Astrometry 270 d}  &
\colhead{Astro+ all RV 50 d}  &
\colhead{Astro+ all RV 270 d}
}
\startdata
$X_0$(mas)              &        -16.6 $ \pm $   1.6 &    -14.1$^d$  $ \pm $   3.2   &   -21.15    $ \pm $    2.3    &         -17.9	$\pm$  4.7   \\
$Y_0$(mas)              &       -408.0 $ \pm $   1.9 &   -406.1$^d$  $ \pm $   3.5   &  409.52     $ \pm $   2.8     &        -410.5	$\pm$  5.51  \\
$\mu_{\rm R.A.}$(mas/yr)&       -588.98$ \pm $  0.25 &  -589.08      $ \pm $  0.25   &  -588.66    $ \pm $   0.29    &        -589.21	$\pm$  0.26  \\
$\mu_{\rm Dec}$(mas/yr) &      -1360.95$ \pm $  0.25 & -1361.08      $ \pm $  0.24   &  -1361.02   $ \pm $    0.25   &       -1361.36	$\pm$  0.20  \\
$\pi$(mas)              &       168.3  $ \pm $  1.51 &    169.5      $ \pm $   1.4   &    169.95   $ \pm $    1.37   &        169.24	$\pm$  1.30  \\
$v_0$(km/s)             &       35.2   $ \pm $ 1.4   &    35.4$^d$   $ \pm $ 1.050   &    36.06    $ \pm $    0.11   &       36.05	$\pm$  0.08  \\
$v_{\rm offset}$(km/s)  &          -                 &         -                     &      1.5    $ \pm $    0.42   &         1.5      $\pm$  0.36  \\
    &&&&                                                                                                                    			     \\
$P$(days)               &      49.7    $ \pm $   0.5 &    272.1      $ \pm $   4.1   &    49.84    $ \pm $    0.11   &   278.5  	$\pm$	2.7  \\
$Mass$(M$_{\rm J}$)     &      17.5    $ \pm $   4.4 &	  7.1	     $ \pm $   2.7   &	  13.7     $ \pm $    6.4    &     5.0  	$\pm$	2.9  \\
$e$                     &  0.22$^{c}$  $ \pm $  0.30 &	 0.48$^{c}$  $ \pm $  0.31   &	   0.91    $ \pm $    0.13   &     0.90         $\pm$	0.16 \\
$i$(deg)                &        93    $ \pm $     5 &	   90	     $ \pm $    15   &	      4    $ \pm $    5      &     110$^{c}$    $\pm$	50   \\
$\Omega$(deg)           &        40    $ \pm $    20 &     220       $ \pm $    25   &   13$^{c}$  $ \pm $    100    &      40$^{c}$    $\pm$	66   \\
$\omega$(deg)           &        20    $ \pm $    40 &	   30$^{c}$  $ \pm $    80   &	 122$^{c}$ $ \pm $    60     &      17$^{c}$  	$\pm$	90   \\
$M_0$(deg)              &       270    $ \pm $     0 &    170$^{c}$  $ \pm $   108   &	 340$^{c}$ $ \pm $    70     &     156$^{c}$    $\pm$	80   \\
                        &                            &                               &                               &  			     \\
a(AU)$^{d}$             &  0.12                      & 0.36                          & 0.12                          &  0.36			     \\
                        &                            &                               &                               &  			     \\
$\bar{\chi}_0^2$        &   2.28                     & 2.28                          & 2.75                          &  2.75			     \\
$\bar{\chi}^2$          &   0.87                     & 0.93                          & 1.62                          &  1.76			     \\
\enddata
\tablenotetext{a}{The mass of VB 10 is assumed to be $0.078$ M$_\odot$ according to \citet{Pravdo:2009}}
\tablenotetext{b}{\,Large uncertainty due to correlation with the eccentricity}
\tablenotetext{c}{Unconstrained or poorly constrained}
\tablenotetext{d}{Derived quantity using Kepler equations}
\end{deluxetable}

\end{document}